\documentclass[11pt]{article}

\usepackage{amssymb}
\usepackage[pdftex]{color,graphicx}
\usepackage[paperwidth=5.3in, paperheight=7.5in, top=.55in, bottom=.5in, left=.14in, right=.14in]{geometry}
\usepackage{textcomp}

\newtheorem{theorem}{Theorem}

\newcommand{\ipar}{\begin{list}{\hbox{$\:$}}{} \item[] }
\newcommand{\rapi}{\end{list}}

\begin{document}

\title{\bf Maximium Priority Matchings}
  
\author{Jonathan Turner}
\date{\normalsize {\sc wucse}-2015-06}
\maketitle

\begin{abstract}
Let $G=(V,E)$ be an undirected graph with $n$ vertices and $m$ edges,
in which each vertex $u$ is assigned an integer priority in $[1,n]$,
with 1 being the ``highest'' priority.
Let $M$ be a matching of $G$. We define the {\sl priority score} of $M$
to be an $n$-ary integer in which the  $i$-th most-significant digit
is the number of vertices with priority $i$ that are incident to an edge in $M$.
We describe a variation of the augmenting path method (Edmonds' algorithm)
that finds a matching with maximum priority score in $O(mn)$ time.
\end{abstract}

\pagestyle{plain}

\section{Introduction}

A {\sl matching} in an undirected graph is a subset of its edges, 
no two of which share a common endpoint.
In the {\sl maximum size matching problem}, the objective is to find a matching
with the largest possible number of edges.
Edmonds showed how to solve the problem for
general graphs~\cite{ed65} and more efficient implementations of his method are
described in~\cite{ga76} and~\cite{mv80}.
Hopcroft and Karp described a simpler algorithm for the case of bipartite graphs~\cite{hk73}.

\begin{figure}[t]
\centerline{\includegraphics[width=3in]{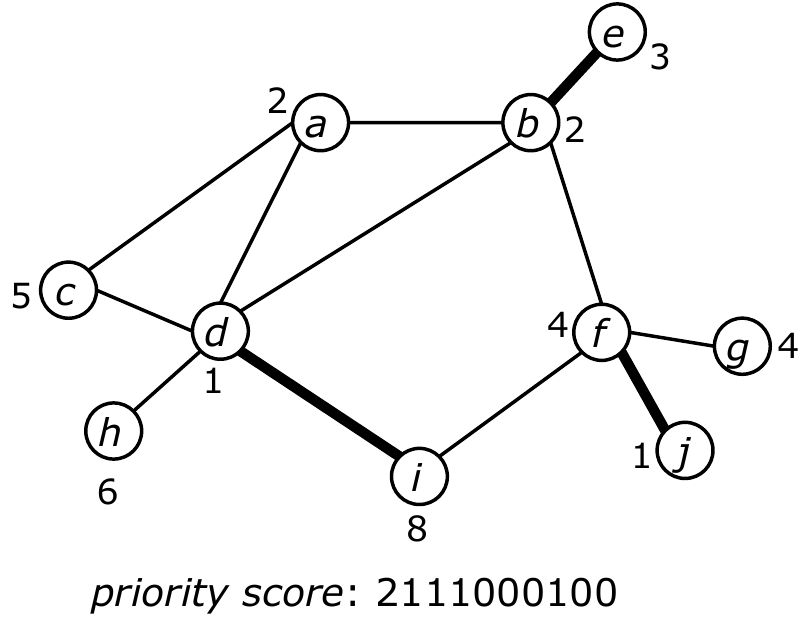}}
\caption{Example showing the priority score for a matching}
\label{priorityMatch}
\end{figure}

A matching is said to {\sl match} a vertex $u$, if one of its edges is
incident to $u$. Given a graph in which vertices are assigned integer priorities $\rho(u)$, 
a {\sl maximum priority matching}
is one that maximizes the number of matched vertices in the highest priority class,
then maximizes the number of matched vertices in the next priority class
(without reducing the number matched in the highest priority class), and so forth.
We define a matching's {\sl priority score} to be the $n$-ary number in which
the $i$-th most-significant digit is the number of matched vertices with priority $i$.
Figure~\ref{priorityMatch} shows an example of a graph with a matching whose
priority score is 2111000100. Adding the edge {\sl ac} yields a matching with
a score of 2211100100. A maximum priority matching is a matching that has
a priority score with maximum value.
This version of the matching problem arises as a subproblem in an approximation
algorithm for an {\sc np}-complete scheduling problem for crossbar switches used
in internet routers~\cite{tu15}.

In this paper, we show how Edmonds' algorithm for the ordinary matching problem
can be extended to solve priority matching problems.
In section 2, we focus on the special case of two priorities ($\rho(u)\leq 2$ for all $u$).
This problem is of independent interest and provides a useful introduction to the
methods used in the more general case, which is discussed in section~3.

\section{Two priority case}

The two priority case can be phrased more simply by defining a set $S$ consisting of
all vertices with priority 1. Our objective is then to find a maximum size matching that
also matches the largest possible number of vertices in $S$.

In Edmonds' algorithm for maximum size matching,
{\sl augmenting paths} are used to convert a matching to a larger one.
Given a matching in a graph, an augmenting path is a simple path in which
the edges alternate between matching edges and non-matching edges, 
with both endpoints unmatched. 
By reversing the status of
the matching and non-matching edges on such a path, we can obtain a new
matching with one more edge than the original. Thus, so long as we can find
an augmenting path, we can expand a given matching. Edmonds showed that
if a larger matching exists, the graph must contain an augmenting path.

For the priority matching problem, we must adjust the definition of an augmenting path.
Let $M$ be a matching of a graph $G$ that does not match all the vertices in the given set $S$.
An {\sl augmenting path} in $G$ with respect to $M$ is a path $p=u_0,\ldots,u_t$
in which every other edge belongs to $M$, $u_0$ is unmatched, $u_0\in S$ and if
$u_t$ is matched, $u_t\not\in S$.

\begin{figure}[h]
\centerline{\includegraphics[width=2.5in]{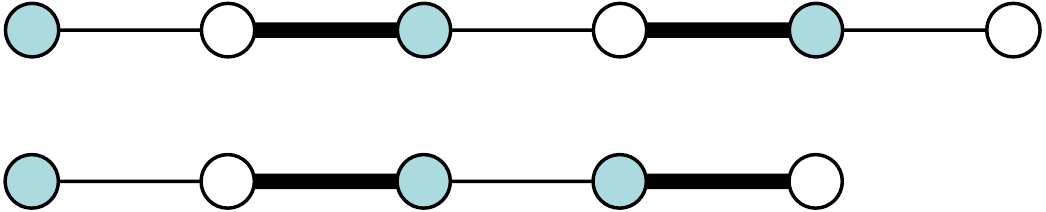}}
\caption{Augmenting paths for two priority case (shaded vertices are in $S$)}
\label{augPath}
\end{figure}

Figure~\ref{augPath} shows examples of two such paths.
Observe that in both cases, if one replaces the path edges in $M$ with the path edges not in $M$,
we get a new matching that matches at least one more vertex in $S$. Also, note that
when we expand a matching in this way, all previously matched vertices in $S$ remain matched.

Define the {\sl count} of a two-priority matching to be the number of vertices in $S$ that are
incident to edges in the matching.
Now consider two matchings $M$ and $M'$, where $M$ has a smaller count than $M'$. 
Let $N$ be the graph defined by the edges that are in $M$ or in $M'$, but not both.
Note that $N$ consists of a collection of disjoint paths and cycles
with alternate edges in $M$ and $M'$. 
Since $M'$ has a larger count than $M$, $N$ must contain at least
one path $p$ in which $M'$ matches more elements of $S$ than does $M$.
Since all interior vertices of a path are matched by both $M$ and $M'$, the
path endpoints must account for the difference in the matchings' counts.
That means that at least one endpoint must be in $S$ and unmatched by $M$.
If the other endpoint is matched, it cannot also be in $S$. 
Hence, $p$ satisfies the condtion for an augmenting path
with respect to $M$. Thus, for any matching that does not have the largest possible count,
there exists an augmenting path that can be used to obtain a new matching with a
larger count.

Observe that given a matching with the maximum possible count, but less than maximum size,
the original augmenting path method can be used to obtain
a matching with the same count, but one more more matching edge.
Hence one can easily convert a maximum count matching to one that has both
maximum count and size. Such a matching also maximizes the number of matched
vertices that are not in $S$, hence it satisfies the definition of a maximum priority matching
for the 2 priority case.

To complete the description of the augmenting path method, we still need an algorithm
to find an augmenting path. Our presentation is an adaptation of that given in~\cite{ta83}
for the maximum size matching problem. We start with the special case of bipartite graphs.

\begin{figure}[t]
\centerline{\includegraphics[width=4.95in]{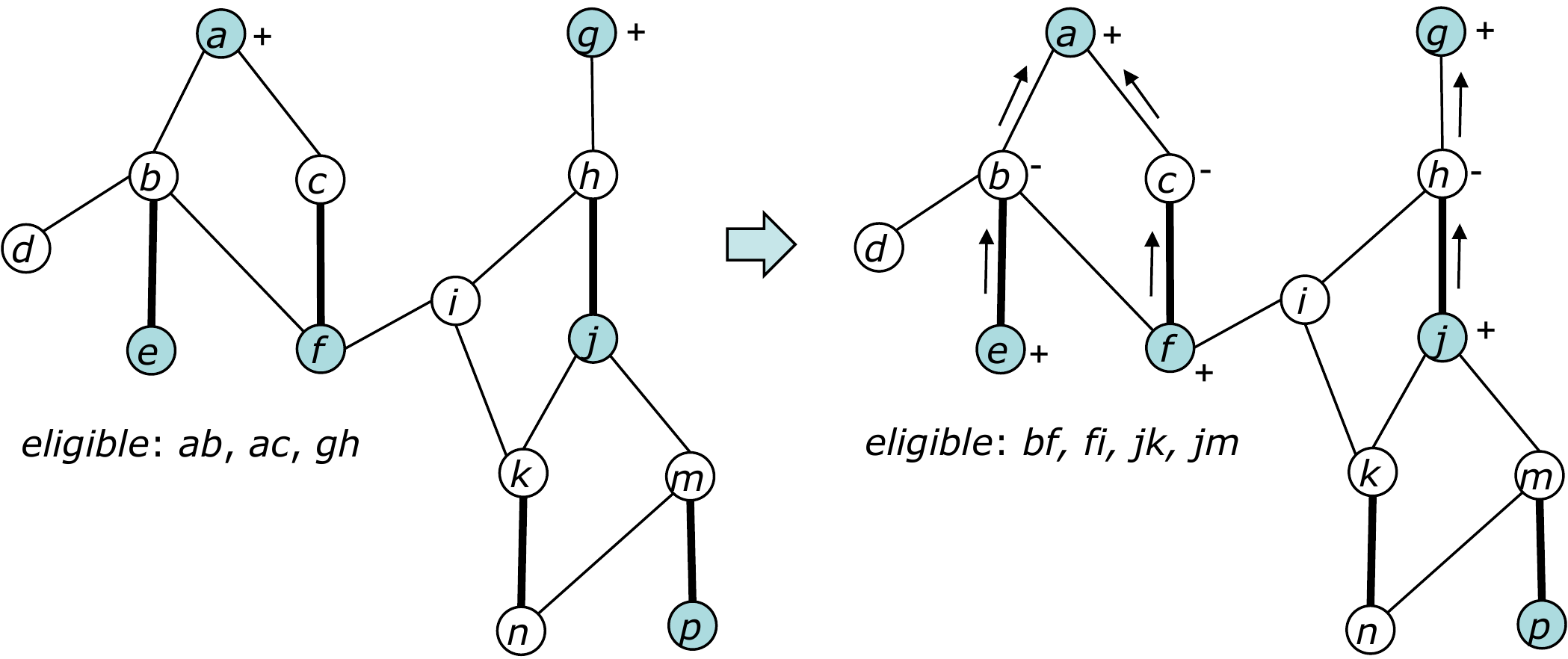}}
\caption{Example of bipartite algorithm at start of augmenting path search and after three steps
(even vertices labelled $+$, odd vertices $-$ and arrows point to parents in trees)}
\label{bipartiteAlgorithm}
\end{figure}

The algorithm finds an augmenting path by building a collection of trees 
rooted at unmatched vertices in $S$. Vertices that have not yet been added to a tree
are called {\sl unreached}, while a vertex $u$ in a tree is called {\sl odd} or {\sl even}
depending on the length of the tree path from $u$ to the root of the tree.
Initially, the unmatched vertices in $S$ are the only tree vertices.
The algorithm also maintains a list of {\sl eligible edges} that initially 
contains all edges incident to tree roots.
It then repeats the following step until it either finds an augmenting path or
runs out of eligible edges.
\ipar
Select an eligible edge $e=\{u,v\}$ for which $u$ is even, remove it from the
eligible list, then apply the applicable case from those listed below.
\begin{itemize}
\item
If $v$ is unreached and matched, let $\{v,w\}$ be the matching edge incident to $v$.
Extend the tree containing $u$ by making $v$ a child of $u$ and $w$ a child of $v$.
If $w$ is not in $S$, then the path from $w$ to the root of its tree is an augmenting path;
otherwise, add all non-matching edges incident to $w$ to the eligible list.
\item
If $v$ is unreached and unmatched then the path consisting of $e$ plus the tree path from $u$
to its tree root is an augmenting path.
\item
If $v$ is even then the path formed by combining $e$ with the tree path from $u$ to the root of 
its tree and the tree path from $v$ to the root of its tree is an augmenting path.
(Note that $u$ and $v$ are in different trees, since the graph is bipartite.)
\item
If $v$ is odd, ignore $e$ and proceed to the next eligible edge.
\end{itemize}
\rapi
Figure~\ref{bipartiteAlgorithm} illustrates the operation of the algorithm.
Note that if edge {\sl bf} is selected next, the algorithm will ignore it, if {\sl fi} is selected next,
it will find the odd-length augmenting path {\sl acfi}, and if edge {\sl jk} is selected next, 
it will find the even-length augmenting path {\sl ghjkn}.

To establish the correctness of the algorithm, we show that if it halts without finding
an augmenting path, then the graph must not contain one. We start by noting a few properties
of the algorithm.
\begin{enumerate}
\item
Every tree root is unmatched and in $S$ and each tree has exactly one unmatched vertex.
\item
Every matching edge either has both endpoints unreached, or one odd and one even.
In the latter case, the even endpoint is the child of the odd endpoint in some tree.
\item
Every edge joining two vertices in the same tree joins an odd vertex to an even vertex.
\item
If the algorithm fails to find an augmenting path, then when it fails, every edge with an endpoint
in some tree has at least one odd endpoint.
\item
If the algorithm fails to find an augmenting path, then when it fails, every even vertex is in $S$.
\end{enumerate}

Properties 3 and 4 imply that if the algorithm fails to find an augmenting,
any path $u_0,u_1,\ldots, u_t$ with $u_1$ a child of $u_0$ that alternates between 
matching and non-matching edges, must also alternate between even and odd vertices.
So in particular, if the algorithm fails to find an augmenting path, but the graph contains 
an augmenting path $p=u_0,\ldots,u_t$ with $u_0$ unmatched and in $S$, 
all vertices in $p$ must alternate between even and odd. 
If $t$ is odd, this implies that $u_t$ is odd and unmatched, but this
contradicts the fact that every odd vertex is matched. If $t$ is even, then $u_t$ is even
and matched, but this contradicts property 5, since the matched endpoint of an augmenting
path cannot be in $S$.

Before proceeding to the case of general graphs,
we show that for the special case where $S$ consists of all
vertices of maximum degree, the algorithm finds a matching that covers all vertices in $S$.
Assume, to the contrary, that the algorithm fails to find an augmenting path when there
is some unmatched vertex in $S$. Consider the collection of trees at the time the algorithm
halts and recall that by property 5, all the even vertices must be in $S$. 
Since every tree has one more even vertex
than it has odd vertices, some even vertex must have an edge that connects it to an unreached
vertex, but this contradicts property 4.
Hence, the algorithm matches all vertices in $S$.

\begin{figure}[t]
\centerline{\includegraphics[width=4.95in]{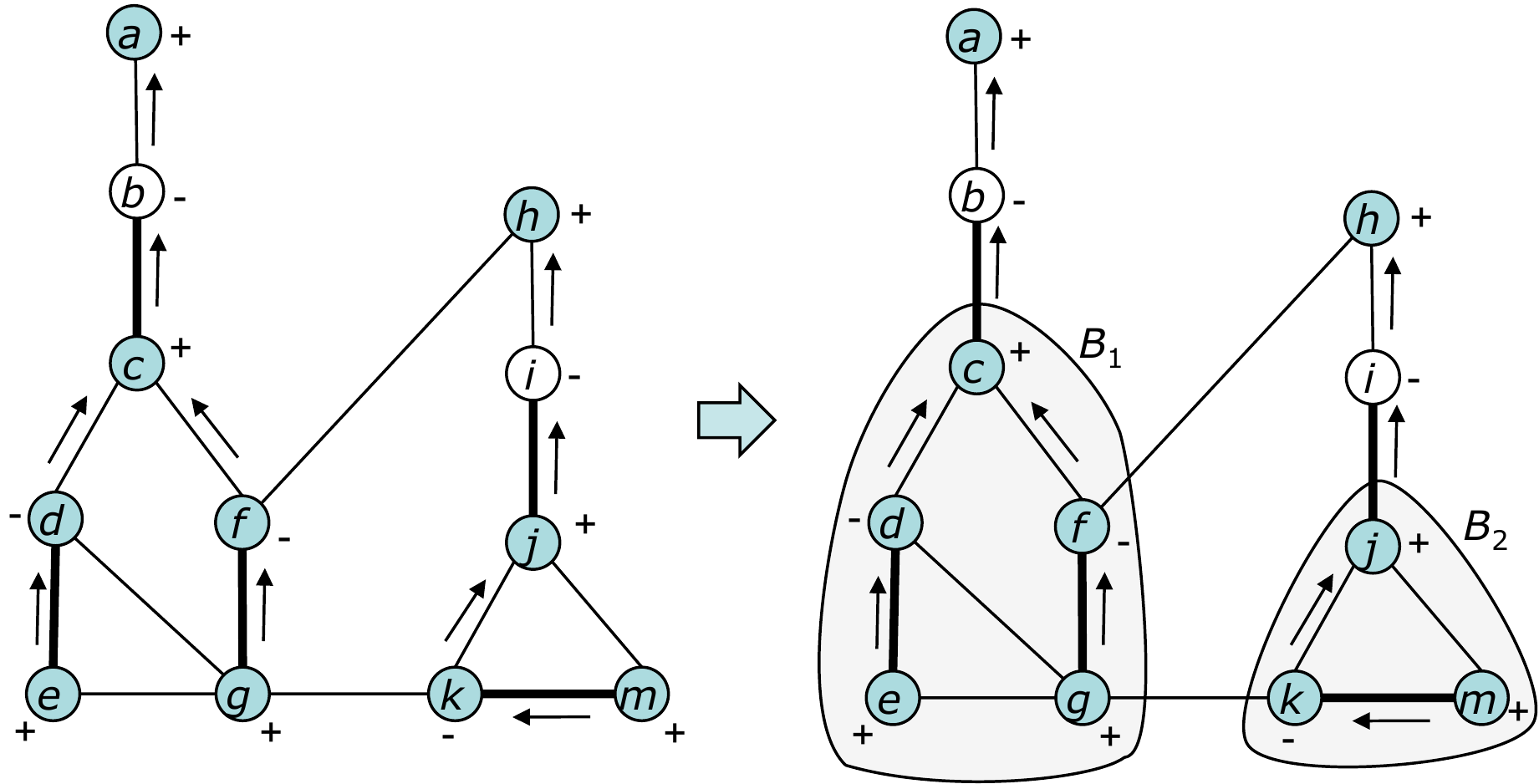}}
\caption{Example of general graph with blossoms}
\label{blossoms}
\end{figure}

Next, we show how to find augmenting paths in general graphs. As with ordinary matchings,
the key issue is handling odd-length cycles, known as {\sl blossoms}. Edmonds showed how to
extend the ordinary augmenting path search to recognize blossoms and {\sl shrink} each
blossom down to a single vertex, producing a new graph which has an augmenting path
if and only if the original graph does. This is illustrated in Figure~\ref{blossoms}, which shows
a blossom $B_1$ containing vertices {\sl c, d, e, f} and {\sl g} and a blossom $B_2$ containing
vertices {\sl j, k} and {\sl m}. In the ``shrunken graph'' on the right, we have two augmenting paths
{\sl abB$_1$h} and {\sl abB$_1$B$_2$ih}. The corresponding paths in the original graph
are {\sl abcdegfh} and {\sl abcfgkmjih}.
There is a straightforward procedure to obtain an augmenting path in the underlying
unshrunken graph, given an augmenting path in the current shrunken graph.

We say than a vertex is {\sl internal} if it is contained in some blossom,
otherwise it is {\sl external}. The {\sl base} of a blossom is the unique vertex
in the blossom that has no incident matching edge with the other endpoint in the blossom.
So, in Figure~\ref{blossoms}, the base of $B_1$ is $c$ and the base of $B_2$ is $j$.

Our algorithm modifies the blossom-shrinking procedure
to accommodate our modified augmenting paths. 
As in the bipartite case, 
the algorithm finds an augmenting path by building a collection of trees 
rooted at unmatched vertices in $S$.
For convenience, we let $\beta(u)$ denote the largest blossom containing a vertex $u$
(if $u$ is external, $\beta(u)=u$).
Initially, all edges incident to tree roots are on the list of eligible edges.
The algorithm then repeats the following step until it either finds an augmenting path 
in the current shrunken graph or runs out of eligible edges.

\ipar
Select an eligible edge $e=\{u,v\}$ for which $\beta(u)$ is even and remove it from the
list of eligible edges.
\begin{itemize}
\item
If $v$ is unreached and matched, let $\{v,w\}$ be the matching edge incident to $v$;
extend the tree by making $v$ a child of $u$ and $w$ a child of $v$.
If $w$ is not in $S$, then the path from $w$ to the root of its tree in the
current shrunken graph is an augmenting path;
otherwise, add all non-matching edges incident to $w$ to the eligible list. 
\item
If $v$ is unreached and unmatched, then the path consisting of $e$ plus the tree path from $\beta(u)$
to the root of its tree in the current shrunken graph is an augmenting path.
\item
If $\beta(v)$ is even and in a different tree than $\beta(u)$,
then the path formed by combining $e$ with the tree path from $\beta(u)$ to the root of 
its tree in the current shrunken graph
and the tree path from $\beta(v)$ to the root of its tree is an augmenting path.
\item
If $\beta(v)$ is even and in the same tree as $\beta(u)$,
then the cycle formed by combining $e$ with the tree paths in the
current shrunken graph from $\beta(u)$ and $\beta(v)$ to their
nearest common ancestor forms a blossom. If some odd vertex $x$ in the cycle is not in $S$,
then there is an augmenting path in the current shrunken graph
from $x$ to the root of its tree that starts with the
matching edge incident to $x$, proceeds around the new blossom to its base and then up
the tree to the root. If there is no such vertex, add all non-tree edges incident to odd vertices
in the blossom cycle to the eligible list and shrink the blossom.
\item
If $\beta(v)$ is odd, ignore $e$ and proceed to the next iteration
\end{itemize}
\rapi
Figure~\ref{blossoms} can be used to illustrate the operation of the algorithm
(note that the shaded vertices are in $S$). When the algorithm reaches the state
shown on the left side of the figure, the eligible edges are {\sl hf, dg, gk, eg} and {\sl jm}.
The algorithm ignores edges {\sl hf, dg} and {\sl gk} and forms one blossom when processing
edge {\sl eg} and the other when processing {\sl jm}. At this point, edges {\sl hf} and
{\sl gk} are again eligible. Processing either edge leads to the discovery of an augmenting path.
Also, observe that if $d$ were not a member of $S$, then when edge {\sl eg} was processed,
the algorithm would have found the augmenting path {\sl abcfged}.
We defer the correctness proof of the algorithm to the next section.

\section{Unlimited priorities}

We can find a maximum priority matching for the general case using a generalization of the 
algorithm for the two priority case. We first maximize the number of priority 1 vertices 
that are matched,
then the number of priority 2 vertices and so forth. At each step, we find an augmenting path
that increases the number of matched priority $i$ vertices without decreasing the number
of priority $j$ vertices, for all $j<i$. 

Define the $i$-score of a matching $M$ as the $n$-ary
integer with $i$ digits in which the $j$-th most significant digit is the number
of priority $j$ vertices that are matched by $M$.
For example, in Figure~\ref{priorityMatch}, the 2-score is 21 and the 5-score is 21110.
Given a matching $M$ with a maximum $(i-1)$-score, an {\sl $i$-augmenting path} is
a path $p=u_0,\ldots,u_t$ in which edges alternate between matching edges and non-matching
edges, $u_0$ is unmatched, $\rho(u_0)=i$, and if $u_t$ is matched, $\rho(u_t)>i$.
Observe that because $M$ has a maximum $(i-1)$-score, if $u_t$ is unmatched, it cannot have
priority less than $i$. Consequently, if we exchange the non-matching edges
and matching edges in $p$, we obtain a new matching with the same $(i-1)$-score and a
larger $i$-score than $M$.

To justify the use of augmenting paths, we must show that if a matching $M$ has
a maximum $(i-1)$-score but not a maximum $i$-score, then there must be an $i$-augmenting path
for $M$. Let $M$ be such a matching and let $M'$ be a matching with a larger $i$-score.
Let $N$ be the graph consisting of edges that are in $M$ or $M'$ but not both.
$N$ consists of a collection of disjoint paths and even-length cycles.
Since $M'$ has a larger $i$-count than $M$, there must be some path $p$ in $N$,
in which $M'$ has a larger $i$-count than $M$. If $p$ is an odd-length path, its endpoints must
be unmatched by $M$ and at least one of its
endpoints must have priority $i$, making $p$ an augmenting path for $M$.
If $p$ is an even length path, the endpoint that is unmatched by $M$ must have priority $i$
and the endpoint matched by $M$ must have priority $>i$. Hence, $p$ is an augmenting
path in this case, as well. We summarize this argument in the following theorem.
\begin{theorem}
Let $G=(V,E)$ be an undirected graph with priorities $\rho(u)$ and let $M$ be a matching with
a maximum $(i-1)$-score. $G$ contains an $i$-augmenting path with respect to $M$ if and
only if $M$ does not have a maximum $i$-score.
\end{theorem}

\noindent To find a matching with maximum overall priority score, we initialize $i=1$, 
then repeat the following step until $i>n$.
\ipar
Search for an $i$-augmenting path; if one is found, augment the matching by
reversing the status of the path edges, otherwise increment $i$.
\rapi
The heart of the method is the algorithm used to find an $i$-augmenting path.
At the start of each path search, all unmatched priority $i$ vertices are tree roots,
and all edges incident to these vertices are in the eligible list.
The algorithm then searches for an $i$-augmenting path
by repeating the following step until it either finds a path or
runs out of eligible edges.
\ipar
Select an eligible edge $e=\{u,v\}$ for which $\beta(u)$ is even and remove it from the
list of eligible edges.
\begin{itemize}
\item
If $v$ is unreached and matched, let $\{v,w\}$ be the matching edge incident to $v$;
extend the tree by making $v$ a child of $u$ and $w$ a child of $v$. 
If $\rho(w)>i$ then the path in the current shrunken graph from $w$ to the root of 
its tree is an $i$-augmenting path;
otherwise, add all non-matching edges incident
to $w$ to the eligible list and continue.
\item
If $v$ is unreached and unmatched, then the path consisting of $e$ plus the tree path in
the current shrunken graph from $\beta(u)$ to the root of its tree is an $i$-augmenting path.
\item
If $\beta(v)$ is even and in a different tree than $\beta(u)$,
then the path formed by combining $e$ with the tree path from $\beta(u)$ to the root of 
its tree in the current shrunken graph and the tree path from $\beta(v)$ to the root of its tree 
is an $i$-augmenting path.
\item
If $\beta(v)$ is even and in the same tree as $\beta(u)$,
then the cycle formed by combining $e$ with the tree paths in the current shrunken
graph from $\beta(u)$ and $\beta(v)$ to their
nearest common ancestor forms a blossom. 
If some odd vertex $x$ on the blossom cycle has $\rho(x)>i$,
then there is an $i$-augmenting path in the current shrunken graph from $x$ to the root of the 
tree that starts with the
matching edge incident to $x$, continues around the blossom cycle to its base and then up
the tree to the root. If there is no such vertex, add all non-tree edges incident to odd vertices
on the blossom cycle to the eligible list and shrink the blossom.
\item
If $\beta(v)$ is odd, ignore $e$ and proceed to the next iteration.
\end{itemize}
\rapi

\noindent
Once again, to establish the correctness of the algorithm, we need to show that if it halts 
without finding an $i$-augmenting path, then the graph must not contain one. 
Note the following properties of the algorithm.
\begin{enumerate}
\item
Every tree root is unmatched and has priority $i$ and each tree has exactly one unmatched vertex.
\item
Every matching edge either has both endpoints unreached, or one odd and one even.
In the latter case, the even endpoint is the child of the odd endpoint in some tree.
\item
For every internal vertex $x$, $\beta(x)$ is even, and if $x$ is unmatched, 
then $\beta(x)$ is unmatched.
\item
If the algorithm fails to find an augmenting path, then when it fails, any edge that has
endpoints that are both even or internal is contained within some blossom.
\item
If the algorithm fails to find an augmenting path, then when it fails, every vertex $x$ 
that is even or internal has $\rho(x)\leq i$.
\end{enumerate}

Now, suppose that the original graph contains an augmenting path $p=u_0,\ldots,u_t$, 
but the algorithm halts without finding a path.
If both endpoints of $p$ are unmatched, then the endpoints cannot be in the same tree
(by property 1). 
In this case, let $\{u_k,u_{k+1}\}$ be an edge in $p$ with $u_{k}$ and $u_{k+1}$
in different trees.
By property 5, at least one of $u_k$ and $u_{k+1}$ must be odd and external.
Assume, without loss of generality, that $u_k$ is odd and external.
Since $\{u_{k-1},u_k\}$ is a matching edge, $k$ is even. 
If just one endpoint ($u_0$) of $p$ is unmatched, then
$t$ is even, $\rho(u_t)>i$ and hence $u_t$ odd and external (by property 4).
Thus, in both cases, $p$ contains an odd external vertex $u_k$ with $k$ even.

Let $j$ be the smallest even integer for which $u_j$ is odd and external.
Since $u_j$ is external, there must some vertex in $\{u_0,\ldots,u_{j-1}\}$ that is
odd and external.
(If not, all vertices in $\{u_0,\ldots,u_{j-1}\}$ must be contained in a common blossom
(property 5), and since $u_0$ is unmatched, the blossom must be also (property 3), hence must include the matching edge $\{u_{j-1},u_j\}$, contradicting the fact that $u_j$ is external.)
Let $i$ be the largest integer $<j$ for which $u_i$ is odd and external
and note that $i$ must be odd. 
This implies that all vertices in $\{u_{i+1},\ldots,u_{j-1}\}$ are in a common blossom
and that blossom is incident to two matching edges $\{u_i,u_{i+1}\}$ and $\{u_{j-1},u_j\}$.
This contradiction implies the correctness of the algorithm.
(This argument was adapted from~\cite{ta83}.)

Each augmenting path search can be implemented to run in $O(m\log n)$ time using the
method described in~\cite{ga76} to represent the current shrunken graph.
This can be reduced to $O(m)$ time using the data structure described in~\cite{gt83}.

\begin{theorem}
The augmenting path algorithm for priority matching computes a matching with maximum 
priority score. It can be implemented to run in $O(mn)$ time.
\end{theorem}

We close by noting that the maximum priority matching is also a maximum size matching.
If it were not, we could find an ordinary augmenting path in the graph that would match
two more vertices, giving it higher priority score.

\section{Closing remarks}

The maximum size matching problem can be solved in $O(mn^{1/2})$ time using the
algorithms described in~\cite{hk73} for the bipartite case and~\cite{mv80} for the general
case. It is possible that one or both of these algorithms could be adapted to handle
maximum priority matching.

It might also be interesting to consider a weighted version of the problem.
In the maximum weight matching problem, we seek a matching that maximizes the
sum of the edge weights of the edges in the matching. While one cannot simultaneously
maximize the weight and priority score of a matching, one could conceivably maximize
the weight of a matching with a specified minimum priority score, or maximize the priority score
of a matching with a specified minimum weight. Alternatively, one might associate weights
with vertices and find matchings that maximize the weight of matched vertices.


\begin{thebibliography}{99}

\bibitem{ed65}
Edmonds, Jack. ``Paths, trees and flowers,'' {\sl Canadian Journal of Mathematics}, 
1965, pp. 449--467.

\bibitem{ga76}
Gabow, Harold N.
``An efficient implementation of Edmonds' algorithm for maximum matching on graphs,''
{\sl Journal of the Association for Computing Machinery}, 1976, pp. 221--234.

\bibitem{gt83}
Gabow, Harold N and Robert E. Tarjan.
``A linear time algorithm for a special case of disjoint set union,''
{\sl ACM Symposium on the Theory of Computing (STOC)}, 1983, pp. 246-251.

\bibitem{hk73}
Hopcroft, John E. and Richard M Karp.
``An $O(n^{5/2}$ algorithm for maximum matching in bipartite graphs,''
{\sl SIAM Journal on Computing}, 1973, pp 225--231.

\bibitem{mv80}
Micali, Silvio. and V. V. Vazirani.
``An $O(\sqrt{|V|}\cdot |E|)$ algorithm for finding maximum matchings in general graphs,''
{\sl IEEE Symposium on the Foundations of Computer Science (FOCS)}, 1980, pp. 17-27.

\bibitem{ta83}
Tarjan, Robert E.
{\sl Data structures and network algorithms}.
Society for Industrial and Applied Mathematics, 1983.

\bibitem{tu15}
Turner, Jonathan S.
``The bounded edge coloring problem and offline crossbar scheduling,''
Washington University Computer Science and Engineering Department technical report,
{\sc wucs-2015-07}, 2015.
\end{thebibliography}
\end{document}